\documentclass[titlepage,twoside,12pt]{article}
\usepackage{amssymb}
\usepackage{amsfonts}
\begin{document}

\pagenumbering{arabic}
\setcounter{page}{1}

{\Large \bf Deficient Mathematical Models of \\ \\ Quantum Theory} \\ \\

{\it Elem\'{e}r E Rosinger \\
Department of Mathematics \\
University of Pretoria \\
Pretoria, 0002 South Africa \\
e-mail : eerosinger@hotmail.com} \\ \\

{\bf Abstract} \\

The lack of the Max Born interpretation of the wave function as a probability density
describing the localization of a quantum system in configuration space is pointed out related
to the recent category based model of quantum mechanics suggested in Abramski \& Coecke [1,2] and
Coecke. \\ \\

All major theories of Physics, including General Relativity, have a rigorous and consistent
overall mathematical model, except for Quantum Theory. \\
Quantum Field Theory, in this regard, excels in deficiency by mostly using a "crazy quilt" of
ad-hoc pieces of disparate and hardly at all rigorous mathematics. What tries to hold these
patched up pieces together, and also on the proper side of sense, is the intuition and
understanding of the respective physicists involved in their use and manipulation. \\
Furthermore, and apparently unknown to most physicists, even the comparatively simpler theory
of non-relativistic finite quantum systems, including the Dirac bra-ket notation, is still
missing a rigorous and consistent overall mathematical model, in spite of the no less than
three mathematical models von Neumann had suggested in the relatively short period of
1928-1936, see Rosinger [1, pp. 9-13]. \\

This situation regarding the deficient mathematical modelling of Quantum Theory leads to the
following possible comments :

\begin{itemize}

\item the need for, and usefulness of a rigorous and consistent overall mathematical model for
any theory of Physics is not a mere matter of pedantry, be in on the part of some physicists
or mathematicians,

\item the exception by the mentioned deficiency which, so far, Quantum Theory alone exhibits
in the whole of Physics, need not be seen as the harbinger of a new and fortunate permanent
paradigm or state of affairs to come,

\item it may well happen that completely new ways of Mathematics and/or Logic would be needed
in order to obtain for Quantum Theory a rigorous and consistent overall mathematical model.

\end{itemize}

In this regard, recently, in Abramski \& Coecke [1,2] and Coecke, a surprisingly novel and
well motivated mathematical model was proposed for Quantum Computation, or more precisely, for
the usual axioms of Quantum Mechanics of non-relativistic finite systems, as formulated
initially by von Neumann, back in the late 1920s. This new model which uses the {\it abstract}
level of {\it compact closed categories with biproducts}, thus at first sight may appear as
far removed from any sort of genuine Physics, proves nevertheless to be well suited even to
the extent that it can recover the Max Born rule according to which the probability of
measuring the eigenvalue $\lambda_i$, which corresponds to the eigenstate $\xi_i$ as the
outcome of a measurement, has the numerical value $|~<~ \psi,~ \xi_i ~> ~|~^2$, where $\psi$
is the state of the quantum system immediately prior to measurement. \\

Here however, we may recall that there is another Max Born rule as well, namely, the Max Born
interpretation of a normalized wave function $\psi$ as a {\it probability amplitude} relevant
to the {\it localization in configuration space} of the quantum system, see Gillespie,
Shankar. Indeed, suppose that the non-relativistic finite quantum system has the configuration
space $\mathbb{R}^d$, and correspondingly, the state space given by the Hilbert space ${\cal
L}^2(\mathbb{R}^d)$. In this case, if the state of the quantum system is given by the
normalized wave function $\psi \in {\cal L}^2(\mathbb{R}^d)$, and if $B \subseteq
\mathbb{R}^d$ is any Borel subset in the configuration space, then \\

$~~~~~~~~~~~~~ \int_B |\, \psi ( x ) \,|\,^2 \,dx $ \\

is supposed to give the probability of finding the quantum system localized in $B$. \\

Here it should be noted that the configuration space, and within it, the Max Born
interpretation of the wave function simply cannot be disregarded or set aside. Indeed, without
them even the computations related to most simple quantum systems cannot be performed, such a
for instance, those concerning one single one dimensional quantum particle considered in
various potentials, involved in scattering, manifesting tunnelling, and so on. \\

In this regard, the still ongoing popularity of von Neumann's first mathematical model of
Quantum Theory, in spite of its shortcomings, see Rosinger [1, pp. 9-13], is to a good extent
due precisely to the explicit, simple and natural presence in it of configuration space, and
of the Max Born interpretation of the wave function. \\
However, the shortcomings of his first model had led von Neumann to the development of his
second and third models, see Rosinger [1, pp. 9-13]. As it happens, both of these models lack
the ability to identify and describe the configuration space, and thus in particular, are
unable to reproduce the Max Born interpretation of the wave function. \\

So far, the new and rather impressive mathematical model in Abramski \& Coecke [1,2] and
Coecke, exhibits the same shortcomings as the second and third von Neumann models by not being
able to identify and describe the configuration space, and thus in particular, is unable to
reproduce the Max Born interpretation of the wave function. \\

Connected with the three mathematical models of Quantum Theory suggested by von Neumann, it is
particularly important to note that, as far back as in 1935, von Neumann made the
declaration, see Rosinger [1, pp. 9-13] :

\begin{quote}

"I would like to make a confession which may seem immoral : I do not believe in Hilbert spaces
anymore."

\end{quote}

Remarkably, and much unlike in the case of a majority of physicists, or for that matter,
mathematicians in Functional Analysis, the authors of the mentioned new mathematical model of
Quantum Theory in Abramski \& Coecke [1,2] and Coecke are fully aware of that rather early
declaration of von Neumann, see Coecke [p. 1].


\begin{thebibliography}{99}

\bibitem{} Abramsky S, Coecke B [1] : A catergorical semantics for quantum protocols.
arXiv:quant-ph/0402130

\bibitem{} Abramsky S, Coecke B [2] : Abstract physical traces. Theory and Applications of
Categories. Vol. 14, No. 6, 2005, 111-124

\bibitem{} Coecke B : Kindergarten quantum mechanics. \\ arXiv:quant-ph/0510032

\bibitem{} Gillespie D T : A Quantum Mechanics Primer. International Textbook, London, 1973

\bibitem{} von Neumann J : Mathematische Grundlagen der Quantum Mechanik. Springer, Berlin,
1932. In translation Mathematical Foundations of Quantum Mechanics. Princeton Univ. Press,
1955

\bibitem{} Rosinger E E [1] : What is wrong with von Neumann's theorem on "no hidden
variables". arXiv:quant-ph/0408191

\bibitem{} Rosinger E E [2] : Where and how does it happen ? \\ arXiv:physics/0505041

\bibitem{} Rosinger E E [3] : What scalars should we use ? \\ arXiv:math.HO/0505336

\bibitem{} Rosinger E E [4] : Notes on quantum mechanics and consciousness.
arXiv:quant-ph/0508100

\bibitem{} Shankar R : Principles of Quantum Mechanics. Plenum Press, New York, 1980

\end{thebibliography}
\end{document}